# Electrical-current-induced magnetic hysteresis in self-assembled vertically aligned La$_{2/3}$Sr$_{1/3}$MnO$_3$:ZnO-nanopillar composites

W. Pan[1], P. Lu[1], J.F. Ihlefeld[1,2,3], S.R. Lee[1], E.S. Choi[4], Y. Jiang[4], Q.X. Jia[5,6]

[1] Sandia National Laboratories, Albuquerque, New Mexico 87185, United States
[2] Department of Materials Science and Engineering, University of Virginia, Charlottesville, Virginia 22904, United States
[3] Department of Electrical and Computer Engineering, University of Virginia, Charlottesville, Virginia 22904, United States
[4] National High Magnetic Field Laboratory, Tallahassee, Florida 32310, United States
[5] Department of Materials Design and Innovation, University at Buffalo - The State University of New York, Buffalo, New York 14260, United States
[6] Department of Physics, Konkuk University, Seoul, 05029, Korea

**Abstract**

Magnetoresistive random-access memory (MRAM) is poised to become a next-generation information storage device. Yet, many materials challenges remain unsolved before it can become a widely used memory storage solution. Among them, an urgent need is to identify a material system that is suitable for downscaling and is compatible with low-power logic applications. Self-assembled, vertically-aligned La$_{2/3}$Sr$_{1/3}$MnO$_3$:ZnO nanocomposites, in which La$_{2/3}$Sr$_{1/3}$MnO$_3$ (LSMO) matrix and ZnO nanopillars form an intertwined structure with coincident-site-matched growth occurring between the LSMO and ZnO vertical interfaces, may offer new MRAM applications by combining their superior electric, magnetic ($B$), and optical properties. In this Rapid Communication, we show the results of electrical current induced magnetic hysteresis in magneto-resistance measurements in these nano-pillar composites. We observe that when the current level is low, for example, 1 µA, the magneto-resistance displays a linear, negative, non-hysteretic $B$ field dependence. Surprisingly, when a large current is used, $I > 10$ µA, a hysteretic behavior is observed when the $B$ field is swept in the up and down directions. This hysteresis weakens as the sample temperature is increased. A possible spin-valve mechanism related to this electrical current induced magnetic hysteresis is proposed and discussed.

Magnetoresistive random access memory (MRAM) [1,2] utilizes the change of electrical resistance in a magnetic ($B$) field to write, read, and store information. It has emerged as a promising candidate to replace conventional semiconductor memory structures, such as static RAM, dynamic RAM and flash memory, which are facing significant challenges of scaling capability as the semiconductor chip industry moves to smaller nodes. Indeed, as the size of these conventional memory structures is decreased, for example down to 20 nm, leakage current presents a power consumption issue. MRAM, being nonvolatile in nature, has the potential to reduce power consumption drastically. At present, the size of MRAM bit cells is larger than that of conventional semiconductor memory devices. Consequently, a large driving current is required to change the associated magnetic states, thus presenting a power consumption issue. Therefore, one of the pressing needs is to identify new MRAM materials that are suitable for nanoscale device implementations and are compatible with low-power logic applications.

In this Rapid Communication, we explore the current induced magnetic properties of heteroepitaxial $(La_{2/3}Sr_{1/3}MnO_3)_{0.5}$:$(ZnO)_{0.5}$ (LSMO:ZnO) nanocomposites [3]. The phenomena observed from our experiments have suggested novel magnetic structures of such nanocomposites for potential MRAMs application. These materials are grown using both pulsed-laser deposition (PLD) and rf magnetron sputtering, structurally characterized using aberration-corrected scanning-transmission electron microscopy (AC-STEM) and x-ray diffraction (XRD), and electronically studied using low-temperature magneto-transport measurements. AC-STEM of the heteroepitaxial nanocomposite films reveals self-assembled, vertically-aligned ZnO nanopillar arrays featuring previously unreported nanometer-scale compositional redistribution around ZnO nanopillars; moreover, transport measurements demonstrate an electrical current induced magnetic-hysteretic behavior of the nanocomposite.

The small nanopillar size, together with this observation of novel current induced magnetic-hysteretic behavior, makes this material system a potentially ideal platform for next-generation MRAM applications.

Fig. 1a shows a schematic diagram of a self-assembled epitaxial nanocomposite thin film, where LSMO matrix and ZnO nanopillars form an intertwined structure with coincident-site-matched or domain-matched growth [4] occurring between the LSMO and ZnO vertical interfaces [5]. Fig. 1b shows a cross sectional STEM high-angle annular dark field (HAADF) image of such a sample. The thickness of the epilayer is 54 nm. It is clearly seen that vertically aligned epitaxial nanocomposite film is achieved on the $(001)_p$-oriented LaAlO$_3$ (LAO) substrate (where the $p$ denotes pseudo-cubic indices), consistent with the schematic drawing shown in Fig. 1a. The average width of the nanopillars (ZnO) is around 10 nm. Fig. 1c shows a reciprocal space mapping XRD pattern of the (002) reflection of the LSMO:ZnO nanocomposite film. The horizontally broadened LSMO peak shape along $Q_x$ in Fig. 1c reflects the presence of both misfit dislocations and the quasi-periodic pillar structures in the composite epilayer, as both phenomena alter the lateral coherence of the epilayer's crystal structure. Additional {103} and {113} reciprocal space maps (not shown) find that the LSMO lattice-mismatch strain at the growth interface to the LAO substrate is ~80% relaxed.

Further XRD measurements of the $(10\bar{1}0)$ and $(11\bar{2}0)$ reflections of the ZnO find the same epitaxial orientation of the ZnO relative to the LSMO and LAO as previously reported by Chen *et al.* for similar growths on SrTiO$_3$ (STO) substrates [5], where a $[11\bar{2}0]$ direction of the ZnO lies normal to the sample surface and [0001] of ZnO lies parallel to the surface and along <110> directions of the LSMO/LAO. As discussed by Chen *et al.*, this orientation gives rise to an approximate coincident-site-type or domain-type lattice matching of the composite heterolayer and the substrate [4]. The nanostructure and strain of the composite is particularly influenced by a close domain matching of the LSMO and ZnO along the surface normal, where the height of five (002) LSMO planes (5×1.9425 = 9.713 angstroms) matches to an adjacent stacking of six $(11\bar{2}0)$ ZnO planes (6×1.6249 = 9.749 angstroms). The resulting domain-size matching at the nanopillar- sidewall interfaces is within ~ 0.4%, hence the preferred $(11\bar{2}0)$ ZnO orientation.

In Fig. 2a, we show the magnetoresistance ($R_{xx}$) as a function of temperature ($T$) at zero $B$ field in a LSMO:ZnO nanocomposite with a film thickness of 200 nm. The lateral size of the specimen is 5 mm × 5 mm. The inset of Fig. 2a shows the schematic drawing of measurement configuration. As the temperature falls from 290 K to 1.4 K, $R_{xx}$ also decreases continuously, indicating a metallic behavior. A broad peak is observed to be forming near $T$ ~ 290 K, indicating the onset of a metal-insulator transition. The observed critical temperature, $T_c$ ~ 290 K, is similar to the value reported in a previous study [5]. Also, consistent with previous work, a stronger magneto-resistance is observed at high temperatures close to $T_c$ [6]. Note here the $B$ field is measured in units of Tesla (T). As shown in Fig. 2c and Fig. 2f, the measured normalized magnetoresistance (MR) at $B$ = 1 T (defined as MR = ($R_{xx}(0)$-$R_{xx}(B)$)/$R_{xx}(0)$) is ~ 7% at $T$ = 277K, compared to ~ 0.7% at 4 K.

The main results of this letter are shown in Figs. 2b-2e where $R_{xx}$ is plotted as a function of $B$ field, with each panel showing a different combination of excitation current at a fixed measurement temperature of $T$ = 4 K. When a small excitation current of 0.1 μA is used (Fig. 2b), $R_{xx}$ displays a negative, linear $B$ field dependence. We note here that the trace is quite noisy, exceeding the resolution of our measurements. Possible physical origins for this excess noise are under investigation. $R_{xx}$ is measured for the $B$ field sweeping from -6 to 6 T (red) and then from 6 to -6 T (black). The two curves show good overlap given the noisiness of the low-current measurements. As the current is increased to 1 µA at fixed $T$ (Fig. 2c), the noise level becomes much smaller and the linear $B$ field dependence also becomes much clearer. Still, there is no well-defined hysteresis for the $B$ field as it sweeps up and down.

Notably, the magnetic field dependence changes dramatically when the excitation current is increased to 10 µA at fixed $T$ (Fig. 2d). Here, a pronounced hysteretic behavior is observed for the $B$ as it sweeps up and down. Indeed, as the $B$ field is swept downward from an initial value of 6 T, $R_{xx}$ increases linearly with the decreasing $B$ field until it reaches a maximal value at $B$ ~ -0.2 T. Just below this field, $R_{xx}$ drops relatively sharply for a small segment of the sweep and then abruptly transitions back to a linear regime where $R_{xx}$ now decreases with increasingly negative $B$. Upon reversing the sweep and coming back from -6 T, $R_{xx}$ first increases linearly with the decreasingly negative $B$ field, but in this reverse sweep, the field reaches its maximal value at $B$ ~ +0.2 T. Finally, just above this field, there is again a relatively sharp drop and then a repeated transition back to a linear regime, where now $R_{xx}$ decreases linearly with increasing $B$. These two overlapping-sweep traces clearly show a "butterfly-shaped" hysteretic behavior, where the magnetoresistance centers at about zero field. The hysteretic behavior becomes stronger as higher currents are used (e.g. 50 µA Fig. 2e).

In a control experiment, magnetoresistance is also measured in a pure $La_{2/3}Sr_{1/3}MO_3$ thin film at $T = 4$ K. A hysteretic behavior is seen in this sample. However, unlike in LSMO:ZnO composite film where hysteresis only occurs at higher electric current, in this pure LSMO sample the hysteresis exists at all current levels measured, from 1 µA to 50 µA. Moreover, the strength of the magnetoresistance hysteresis shows nearly no current dependence. These results clearly demonstrate that electronic transport across the interfaces of two domains is responsible for the current-induced hysteresis.

Examining Fig. 2e carefully, we also notice that a very weak local minimum seems to be developing at $B = 0$ T. Indeed, as shown later in Fig. 4a, a double-peak structure (one weak and one strong) is developed around $B = 0$ T and $\Delta R_{xx}$ approaches zero after the linear background is subtracted. Moreover, Fig. 2e shows that there is no drift in the resistance states. Indeed, after the $B$ field is swept from – 1.5 T to 1.5 T, the $B$ field sweeping direction is reversed immediately and swept back to zero for a second time (unlike in Figs. 2b-2d). The second down-sweeping curve (blue) overlaps exactly with the initial downward-sweeping-$B$ curve (black) from 3 T to 0 T. The hysteretic behavior completely disappears at elevated temperatures. In Fig. 2f, we show an $R_{xx}$ curve taken at 277 K, close to $T_c$, where the excitation current is in a high-current regime, 10 µA. Unlike the low-temperature high-current measurements, the magnetoresistance curves for the two $B$ sweeping directions now exactly overlap with each other.

Next, we present results from magnetic torque measurements, which show that the observed hysteretic behavior is most likely related to the magnetization ($M$) of the specimen. In Fig. 3a, a schematic diagram of the capacitive cantilever for magnetic torque measurement is shown in which $B$ is applied at a tilt angle of $\theta$ to the sample normal, and a magnetic torque $\tau = M \times B$ is generated and deflects the cantilever. This deflection is detected by measuring the capacitance ($C$) between the cantilevered metal plate (the top golden rectangle shown schematically in Fig. 3a) where the sample sits and the fixed metal plate (the bottom golden rectangle in Fig. 3a) [7]. As long as the change of the capacitance is small (in our case, less than 1%), the torque is proportional to the measured capacitance minus the zero-field balanced capacitor value $C_0$ (~ 1 pF for our sample) [7]. Consequently, the magnetization can be deduced.

Fig. 3b shows the normalized capacitance value ($C/C_0$) as a function of the magnetic field for magnetometry performed at low temperature ($T = 5$ K). $\theta$ is about 11 degrees. The $B$ field is first swept from 2.5 to -2.5 T. At higher $|B|$ fields, $C/C_0$ shows weak $B$ field dependence. As $B$ is decreased, $C/C_0$ decreases and then displays a weak local maximum at $B \sim + 0.3$ T. A larger peak is observed at $B \sim - 0.2$ T. Sweeping the $B$ field from -2.5 to 2.5 T, a weak peak is observed at $B \sim - 0.3$ T and a larger peak at $B \sim + 0.2$ T. We point out that the positions of the large peaks are the same as the $B$ field positions where the resistance maxima occur in Figs. 2d and 2e when the $B$ field is swept along the same direction. The coincidence can be seen more clearly in Fig. 4 by comparing $C/C_0$ and $\Delta R_{xx}$ of I = 50 µA. Here, $\Delta R_{xx}$ is obtained after subtracting the linear background. This strongly suggests that the hysteretic resistance behavior is related to the magnetization of the specimen. Finally, the $B$ field positions ($\sim \pm 0.3$ T) of the weaker peaks in $C/C_0$ correspond well with the abrupt up-kinks in the magneto-resistance also seen in Fig. 2, and the dips around $B = 0$ T in $C/C_0$ seem to correspond well to the weak peaks in $\Delta R_{xx}$.

Elevated-temperature magnetic torque measurements were also carried out. Fig. 3c shows results for $T = 100$ K. At this temperature, the peaks at $B \sim \pm 0.2$ T become much weaker. After the temperature is raised to $T = 200$ K (Fig. 3d), the peaks at $B \sim \pm 0.2$ T are all but gone, which is consistent with the 277 K disappearance of the magnetoresistance hysteresis seen in Fig. 2f.

The $B$ field dependence of our magneto-transport data suggests that there are two contributions to the magnetoresistance. At higher magnetic field, the intrinsic negative linear magnetoresistance is most likely due to the suppression of spin fluctuations, as observed in Ref. [8]. At a lower magnetic field of around $B = 0$, the magnetoresistance hysteresis appears to be related to an electronic transport process across the interfaces between two domains, which is only active at higher electric field (or higher electrical current). In the following, we will discuss two possible mechanisms that may be responsible for the observed electrical current induced magnetization behavior [9]. First, we note that the "butterfly" shape of the magnetoresistance hysteresis is very similar to that in a spin valve device [10-12], where the hysteresis is due to the switching of magnetization in the so-called free layer. Based on this similarity, it is natural to speculate that the same spin valve mechanism is also responsible for the hysteresis observed in our experiment.

Our high-resolution STEM analysis supports this hypothesis. In Fig. 5a, we show the STEM-HAADF image in plan-view of an LSMO:ZnO nanopillar composite grown by magnetron sputtering method (details of growth are given in the methods section). A transition region is visible around ZnO in this HAADF image. To further examine this transition region, STEM energy-dispersive x-ray spectroscopy (EDS) mapping was carried out for the sample. Fig. 5b shows a composite color EDS map, made of zinc $K$ (red), strontium $K$ (blue)), lanthanum $L$ (green), and manganese $K$ (cyan) maps. The individual map for each element is also shown in Fig. 5c. A core-shell structure around the ZnO nanopillars, corresponding to a (La, Mn)-rich and Sr-deficient region, is clearly visible in Fig. 5b. This compositional nanostructure is uniquely enabled by the spontaneously formed nanocomposite architecture.

Due to the observed Sr-deficiency, it can be expected that the shell layer may possess a different magnetization, compared to the nominal, stoichiometric $La_{2/3}Sr_{1/3}MnO_3$ composition. When a low current is used, this Sr-deficient shell layer maintains a ferromagnetic ground state as in $La_{2/3}Sr_{1/3}MnO_3$. Consequently, no magnetization switching, and thus hysteresis, is expected. The resulting low-current electronic transport is likely due to spin polarized electron transport between the LSMO and the Sr-poor LSMO domains, which leads to a linear $B$ field dependence, as was previously observed in Ref. [8] for structurally analogous inter-grain transport for polycrystalline LSMO. On the other hand, a high electric field (or a high electrical current) can modify the magnetic anisotropy and thus alter the direction of magnetization [13-15], and an electric field (or electrical current) induced transition from ferromagnetic to antiferromagnetic ground states may occur [14]. In this regard, it is possible that at higher electrical currents in our experiments, the magnetization of the Sr-deficient shell layer is tilted from (or in the extreme limit antiparallel to) that of LSMO. In this configuration, a spin valve behavior is expected and this can give rise to the observed $B$ field induced resistance hysteretic behavior.

An alternative possibility is related to the possible magnetization of the ZnO nanopillars themselves, not the Sr-poor layer. It is known that a small amount of background carbon may incorporate into the film during the composite-film growth [3]. As a result, the ZnO nanopillars may become weakly ferromagnetic [16]. In addition, oxygen vacancies in ZnO can also induce ferromagnetism [17]. As already noted above, a high electric field (or electrical current) can vary the direction of $M$ [13-15]. In this alternative scenario, it is possible that at a low electrical current, the $M$ in ZnO is aligned with the $M$ in LSMO, and electronic transport is again dominated by spin polarized electron transport, but now between the LSMO and ZnO nanopillars (instead of the between the LSMO and the Sr-poor LSMO domains as described above), and a negative linear $B$ field dependence is again expected [8]. At a high electrical current, the $M$ in ZnO may lie in the plane, and as the amplitude of $|B|$ field increases, $M$ eventually switches its direction and becomes parallel to the $M$ in LSMO. This switching behavior of the $M$, now occurring in the ZnO, can again give rise to a butterfly-shaped magnetoresistance hysteresis, as observed.

A few remarks are in order. First, we note that the current density to induce the magnetic hysteresis is quite low in our sample. The cross-sectional area of the contact is about 200 nm × 1 mm, and at the observed critical current between approximately 1 and 10 μA, the critical current density becomes only ~ $5 \times 10^3$ - $5 \times 10^4$ A/m$^2$, significantly lower than ~ $10^6$ A/m$^2$ in a typical MRAM device [9]. This small current density may find unique applications in low-power memory devices. Second, we point out that the electrical-current-induced magnetic hysteresis is observed in a planar device structure, which differs from the sandwich-like device structure typical of currently used MRAM memory cells. The planar-device alternative would enable increased magnetic-memory-cell density, and would also make the cell-fabrication process much easier. Third, among the many available composite heterostructures formed using LSMO [3], the LSMO:ZnO composite forms one of the most promising self-organized structures. This promise arises, in part, from the ability to tune the heteroepitaxial strain state, the compositional distribution, the nanostructure, and the resultant magneto-electronic properties of the LSMO:ZnO composite. As shown in previous work [3], the tunability of strain and structure specifically arises through the introduction of differing composite-layer stoichiometries and the use of differing substrates (LAO, STO, and others). As we have shown here, the attendant variations in structure directly alter the magneto-electronic properties. Beyond these nanostructural effects, La$_{2/3}$Sr$_{1/3}$MnO$_3$ is a *p*-type, metallic ferromagnet, whose magnetic properties can be readily tuned by dopant-induced control of the carrier concentration. We also note that ZnO is an *n*-type large band-gap semiconductor; it is earth abundant and non-toxic; and its electrical properties can be tuned by opto-electrical means [18]. Thus, in heteroepitaxially fabricated LSMO:ZnO *p-n* junctions, one should be able to control and manipulate the electrical and magnetic properties by a variety of magneto-opto-electrical means.

Such devices may ultimately enable a new generation of opto-magneto MRAM applications. Of course, all the measurements reported in this manuscript were carried out at cryogenic temperatures, and thus, LSMO:ZnO composites remain impractical for immediate, real-world applications. Nonetheless, as these promising materials and their physics become more thoroughly explored, understood, and optimized, one may be able to realize the electrical current induced magnetic hysteresis at substantially higher temperatures, perhaps even room temperature.

This work was supported by a Laboratory Directed Research and Development project at Sandia National Labs. Sandia National Laboratories is a multimission laboratory managed and operated by National Technology and Engineering Solutions of Sandia LLC, a wholly owned subsidiary of Honeywell International Inc., for the U.S. Department of Energy's National Nuclear Security Administration under contract DE-NA0003525. This work was performed, in part, at the Center for Integrated Nanotechnologies, an Office of Science User Facility operated for the U.S. Department of Energy (DOE) Office of Science by Los Alamos National Laboratory (Contract DE-AC52-06NA25396) and Sandia National Laboratories (Contract DE-NA-0003525). A portion of this work was performed at the National High Magnetic Field Laboratory, which is supported by National Science Foundation Cooperative Agreement No. DMR-1157490 and the State of Florida.

**Figures and Figure Captions:**

Figure 1:

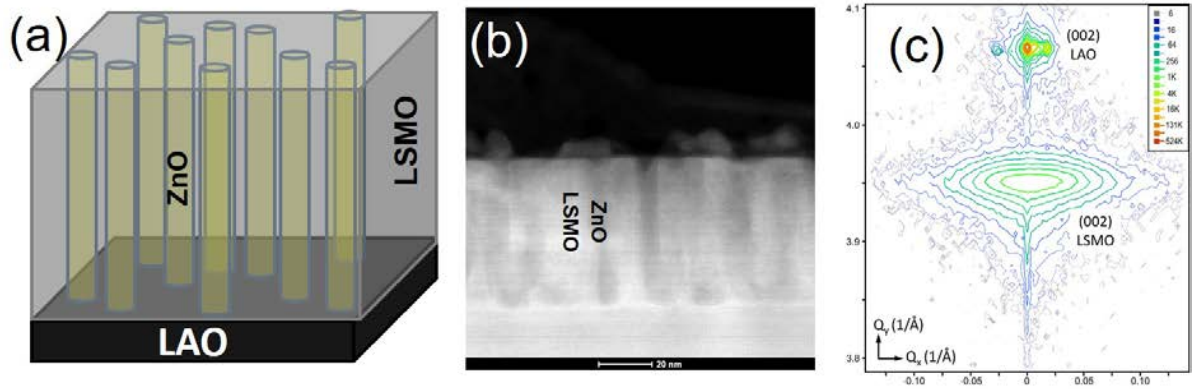

**Figure 1. Structure characterization of self-assembled vertically aligned LSMO:ZnO nanopillars composites**. (a) Schematic diagram of a nanopillars composite sample. (b) High resolution cross-section TEM image. The scale bar is 20 nm. (c) X-ray diffraction (XRD) pattern in the reciprocal lattice. LSMO and LAO peaks are marked.

Figure 2:

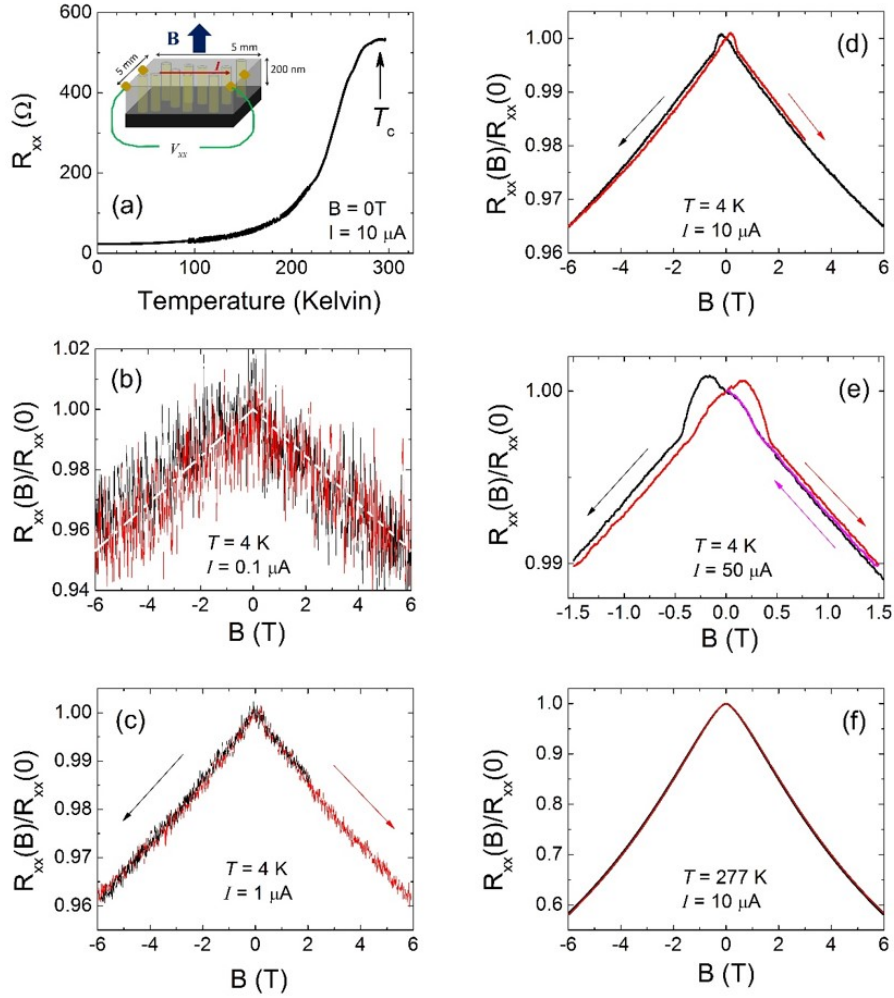

**Figure 2. Magnetoresistance hysteresis of an LSMO:ZnO nanopillars composite sample**. (a) Temperature dependence of the resistance at zero magnetic field. The inset shows the schematic diagram of resistance measurement configuration and dimensions of the specimen. (b)-(e) Normalized magnetoresistance $R_{xx}(B)/R_{xx}(0)$ as a function of $B$ field, with each panel showing a different combination of excitation current $I$ at a fixed measurement temperature of $T$ = 4 K. No hysteresis is observed at $I$ = 0.1 µA (b) and 1 µA (c), while a butterfly-shaped magnetoresistance hysteresis around $B$ = 0 T is observed at $I$ = 10 µA (d) and 50 µA (e). (f) $R_{xx}(B)/R_{xx}(0)$ measured at $T$ = 277K. The excitation current is 10 µA. No hysteresis is observed at this elevated temperature.

Figure 3:

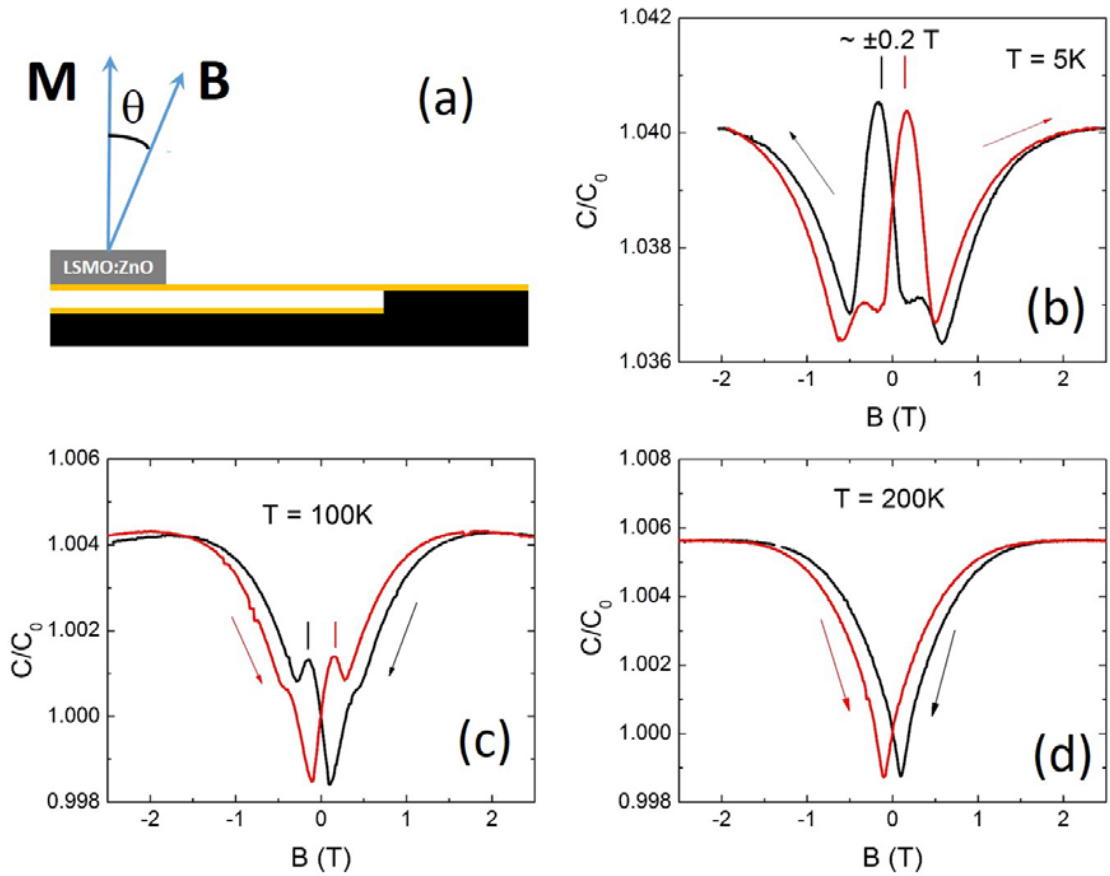

**Figure 3. Magnetic torque measurements**. (a) Schematic diagram of the magnetic torque measurement configuration. θ is the tilt angle between the sample normal and the magnetic (*B*) field. (b) Normalized capacitance *C/C*$_0$ as a function of the *B* field for magnetometry performed at low temperature (*T* = 5 K). The arrows mark the *B* field sweep directions. The vertical lines mark the positions of the two large peaks at B ~ ± 0.2T, respectively. (c) *C/C*$_0$ at *T* = 100 K. (d) *C/C*$_0$ at *T* = 200 K.

Figure 4:

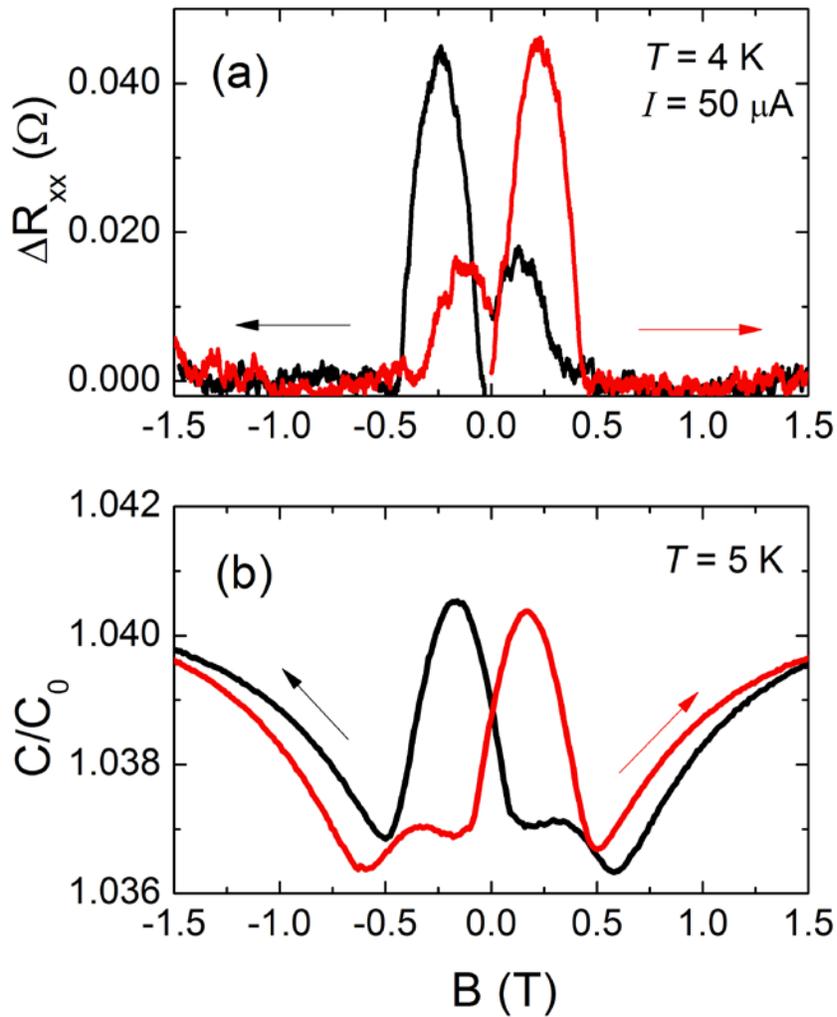

**Figure 4. Comparison of magnetoresistance and magnetic torque measurements**. The positions of the large peaks in the magnetic torque measurement (b) are the same as the *B* field positions where the maxima occur in $\Delta R_{xx}$ (a) when the *B* field is swept along the same direction. $\Delta R_{xx}$ is obtained after subtracting the linear background. This coincidence strongly suggests that the hysteretic resistance behavior is related to the magnetization of the specimen.

Fig. 5

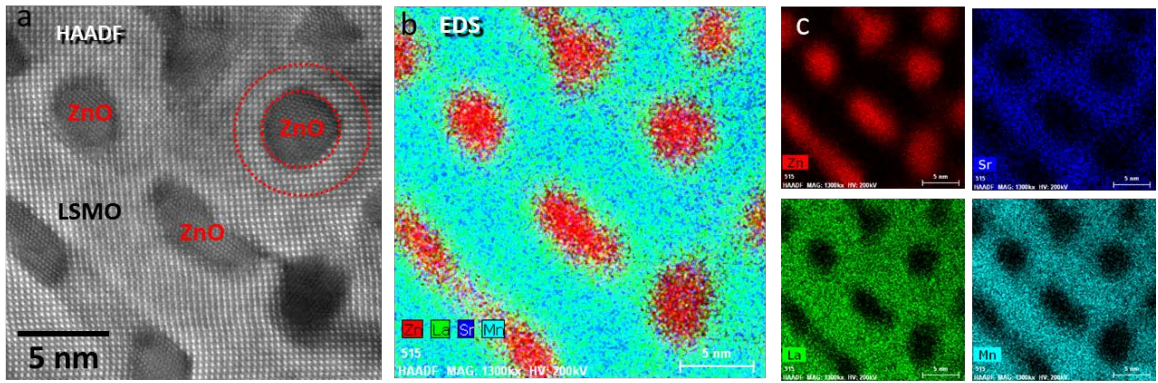

**Figure 5. STEM analysis of an LSMO:ZnO nanopillars composite.** (a) shows an STEM-HAADF image in plan-view. A transition region around ZnO, marked by the red circles, is visible. (b) shows a composite color EDS map, made of zinc *K* (red), strontium *K* (blue), lanthanum *L* (green), and manganese *K* (cyan) maps. A core-shell structure around the ZnO nanopillars, corresponding to a (La, Mn)-rich and Sr-deficient region, is clearly visible. (c) shows the individual map for the above elements.